\documentclass[twocolumn,aps,prd,reprint,nofootinbib,amsmath,floats,floatfix]{revtex4-2}
\usepackage[top=2.5cm, bottom=2.5cm, left=3cm, right=2cm]{geometry}
\usepackage{amsmath,amssymb,mathtools}
\usepackage[normalem]{ulem}
\usepackage[dvipsnames]{xcolor}
\usepackage{graphicx}
\usepackage{enumitem}
\usepackage{hyperref}

\def\be{\begin{equation}}
\def\ee{\end{equation}}
\def\ba{\begin{eqnarray}}
\def\ea{\end{eqnarray}}

\begin{document}

\title{No-go theorem for spontaneous vectorization}

\author{Hsu-Wen Chiang}
\email{jiangxw[at]sustech.edu.cn}
\email{b98202036[at]ntu.edu.tw}

\author{Sebastian Garcia-Saenz}
\email{sgarciasaenz[at]sustech.edu.cn}

\author{Aofei Sang}
\email{12331027[at]mail.sustech.edu.cn}

\affiliation{Department of Physics, Southern University of Science and Technology, Shenzhen 518055, China}


\begin{abstract}
Generalized vector-tensor theories of gravity have drawn attention for admitting hairy black hole solutions, thereby circumventing the standard no-hair theorems. It remains an open question, however, how such black holes may form starting from reasonable initial conditions. It has been suggested that vector hair may grow spontaneously as a result of the field developing a negative effective mass squared --- the so-called spontaneous vectorization mechanism. We demonstrate that this is not possible if the initial state is a hairless black hole, a result that applies to essentially all stationary and axisymmetric solutions of interest in general relativity. More precisely, we prove that the appearance of a negative effective mass squared for the vector field must necessarily be accompanied by ghost- or gradient-type instabilities. Demanding the absence of such instabilities translates into interesting bounds on the coupling constants of the theory as functions of the black hole parameters. In particular, we discover that a Kerr black hole may become unstable when the spin increases above a certain critical value.
\end{abstract}

\maketitle

\section{Introduction} \label{sec:intro}

The classic no-hair theorems~\cite{Bekenstein:1971hc,Bekenstein:1972ky,Bekenstein:1995un,Sotiriou:2011dz} of general relativity (GR) make black holes very promising laboratories to search for new physics. This is so because circumventing the theorems requires some sort of generalization of the paradigm of GR interacting with standard matter. A common avenue in this direction is to envisage additional fields that couple to gravity non-minimally, i.e.\ through curvature couplings. For instance, in the context of scalar-tensor theories, the coupling of a scalar field with the Gauss-Bonnet invariant provides a celebrated example~\cite{Sotiriou:2013qea,Sotiriou:2014pfa,Antoniou:2017acq}.

Once we have a theory that does admit hairy black hole solutions, it is then natural to ask how such black holes may form, say in a realistic astrophysical setting. We can sharpen this question by assuming the theory admits both hairless and hairy states, so that the problem becomes to determine if and when the hairless or GR solution will transition into the hairy one. For scalar-tensor theories, the mechanism of spontaneous scalarization~\cite{Damour:1993hw,Herdeiro:2018wub} provides a well-understood example of such process.

From a fundamental perspective, it appears hard to justify the consideration of a single scalar field in regards to the question of black hole hair, although scalar-tensor theories have undoubtedly proved useful proxies for gaining insight into the issue. It therefore seems sensible to go one step beyond this familiar paradigm and inquire about black hole hair in the form of a vector field. Vector bosons beyond the Standard Model appear as generic predictions in a number of extensions~\cite{Kors:2004dx,Abel:2008ai,Goodsell:2009xc}, and have been widely studied in connection with dark matter~\cite{Arkani-Hamed:2008kxc,Nelson:2011sf,Arias:2012az} and dark energy~\cite{DeFelice:2016yws,deFelice:2017paw,Heisenberg:2020xak,DeFelice:2020sdq,deRham:2021efp,Chiang:2025hrj}. More to the point, consistent vector-tensor theories of gravity have been constructed that accommodate hairy black hole solutions~\cite{Chagoya:2016aar,Minamitsuji:2016ydr,Cisterna:2016nwq,Babichev:2017rti,Heisenberg:2017hwb,Heisenberg:2017xda,Kase:2018owh,Chen:2025aom,Kleihaus:2026rev,Fernandes:2026rjs}.

Analogously to the scalarization effect, the so-called spontaneous \textit{vectorization} mechanism has been proposed as a pathway to the generation of a vector condensate~\cite{Ramazanoglu:2017xbl,Ramazanoglu:2018tig,Annulli:2019fzq,Minamitsuji:2020pak}. In both settings, the field exhibits an instability about the GR solution, driving the growth of hair until an equilibrium state is reached. However, later works unveiled an essential difference between the two mechanisms. Namely, the vectorization instability was found to be driven by ghost- or gradient-type modes, rendering the set-up inconsistent~\cite{Garcia-Saenz:2021uyv,Silva:2021jya,Demirboga:2021nrc,Pizzuti:2023eyt,Hod:2025qip,Chiang:2025gpa} (see also~\cite{Clough:2022ygm,Coates:2022nif,Coates:2022qia,Aoki:2022woy,Unluturk:2023qgk,Rubio:2024ryv}).

On the other hand, these analyses were restricted to some particular static and spherically symmetric backgrounds, leaving open the question of whether more general spacetimes may exist in which vectorization could occur under theoretical control. We provide strong evidence that the answer is negative, by demonstrating that the unwanted instabilities persist for a large class of stationary and axisymmetric metrics, including in particular the Kerr-Newman black hole solution.

\section{Non-minimal gravitational couplings for a vector field} \label{sec:set-up}

We are interested in the conditions of linear stability of a massive vector field $A_{\mu}$ about a GR background, defined here very broadly as any solution for the metric tensor $g_{\mu\nu}$ with vanishing $A_{\mu}$. It has been proved~\cite{Horndeski:1976gi,Garcia-Saenz:2022wsl} that, under some mild assumptions, the unique theory relevant for this question is given by the Lagrangian density
\be \label{eq:lagrangian}
\mathcal{L}=F_{\mu\nu}\mathcal{G}^{\mu\nu\rho\sigma}F_{\rho\sigma}-A_{\mu}\mathcal V^{\mu\nu}A_{\nu} \,,
\ee
where $F_{\mu\nu}\equiv \partial_{\mu}A_{\nu}-\partial_{\nu}A_{\mu}$ and
\be\begin{aligned}
\mathcal{G}^{\mu\nu\rho\sigma}&\equiv - \frac{1}{4}g^{\mu\rho}g^{\nu\sigma} + \alpha\bigg[C^{\mu\nu\rho\sigma} \\
&\quad -\left(g^{\mu\rho}g^{\nu\lambda} - g^{\mu\lambda}g^{\nu\rho}\right) \left(G^\sigma{}_\lambda -\frac{1}{6} \delta^\sigma{}_\lambda G^\tau{}_\tau\right) \bigg] \,, \\
\mathcal V^{\mu\nu}&\equiv \frac{1}{2}\mu^2 g^{\mu\nu}-\beta G^{\mu\nu} \,,
\end{aligned}\ee
in terms of the Proca mass $\mu$, non-minimal coupling constants $\alpha$ and $\beta$, the Einstein tensor $G_{\mu\nu}$ and the Weyl tensor $C_{\mu\nu\rho\sigma}$. The particular structure of $\mathcal{L}$ follows from the requirement that the theory must contain no degrees of freedom beyond those of gravity and a massive spin-1 particle. Indeed, \eqref{eq:lagrangian} can be also obtained from a `top-down' approach by linearizing, about an arbitrary fixed background metric, any non-linear vector-tensor theory, including the well-known Generalized Proca class~\cite{Tasinato:2014eka,Heisenberg:2014rta}.

Our aim is to study the stability of this system in the eikonal approximation. That is, we assume the distance and time scales of variation of the background metric are much greater than the typical wavelength and frequency of the vector field perturbation. As a result, we may assess stability by scrutinizing, at any point in spacetime, the dispersion relation $\omega(k)$ defining the frequency as function of wavenumber $k$.

Although gradient- and tachyon-type instabilities (characterized by complex $\omega$) may be easily diagnosed in this way, the definition of ghost-type instabilities is more subtle, as this requires the existence of a universal time direction along which the Hamiltonian of the system can be defined (see e.g.~\cite{Babichev:2018uiw}). In this event, one can show that a frame exists in which the dispersion relations contain only even powers of $\omega$, and vice-versa, allowing for an unambiguous definition of physical modes.

In this paper, we restrict our attention to the Petrov type D class of spacetimes (defined by the property of having two double principal null directions). In the Newman-Penrose (NP) null basis $\{\ell,n,m,\bar{m}\}$~\cite{Newman:1961qr}, this condition implies that $C_{\mathbf{1212}} = C_{\mathbf{3434}} = 2\,{\rm Re}\, C_{\mathbf{1324}}$, $C_{\mathbf{1234}} = 2i\,{\rm Im}\,C_{\mathbf{1324}}$, with other components of the Weyl tensor zero.\footnote{We use boldface numbers $\{\mathbf{1,2,3,4}\}$ to denote the inner product with $\{\ell,n,m,\bar{m}\}$, respectively. Note also that the metric components read $g_{\mathbf{34}}=-g_{\mathbf{12}}=1$ in our conventions.} Furthermore, we shall assume that only $G^{\mathbf{12}}$ and $G^{\mathbf{34}}$ are non-zero in this basis (notice that these are scalars, hence the assumption is an invariant statement); equivalently, $G^{\mu}{}_\nu$ is diagonal in the NP basis. As we explain below, this set-up includes essentially all stationary and axisymmetric solutions of interest in GR.

It follows that $\mathcal{G}^{\mu\nu\rho\sigma}$ has four independent components and $\mathcal V^{\mu\nu}$ has two independent components. We label these as
\be\begin{aligned} \label{eq:F defs}
F_1&\equiv 4 \mathcal G^{\mathbf{1212}} \,,\\
F_2&\equiv 4 \mathcal G^{\mathbf{3434}} \,,\\
F_3&\equiv - 4\,{\rm Re}\,\mathcal G^{\mathbf{1324}} \,,\\
F_4&\equiv 12\,{\rm Im}\,\mathcal G^{\mathbf{1324}} \,,\\
X_1&\equiv - \mathcal V^{\mathbf{12}}\,,\\
X_2&\equiv   \mathcal V^{\mathbf{34}}\,.
\end{aligned}\ee
This structure guarantees that $\mathcal{G}^{\mu\nu\rho\sigma}$ has no off-diagonal terms in the time-like direction, thus leading to a purely quadratic dispersion relation, as will be made explicit below. We remark that the assumptions made here (Petrov type D spacetime and diagonal $G^{\mu}{}_\nu$ in the NP basis) are sufficient for maintaining a universal time direction. Whether they are also necessary conditions remains an open question.

\section{Mode decomposition}

Given the symmetries of the class of spacetimes under consideration, we expand the vector field locally in a spherical-like orthonormal basis as $A_\mu = \sum_n C_n Z^{(n)}_\mu e^{iS}$, where $S$ is the eikonal phase with $k_\mu = \partial_\mu S$, and
\be\begin{aligned}\label{eq:Z_basis}
Z^{(0)}_\mu &=  \delta_\mu^0  \,,\\
Z^{(1)}_\mu &=  \delta_\mu^1  \,,\\
Z^{(2)}_\mu &=  \ell^{-1} ( k_2 \delta_\mu^2 + k_3 \delta_\mu^3)  \,,\\
Z^{(3)}_\mu &=  \ell^{-1} ( k_3 \delta_\mu^2 - k_2 \delta_\mu^3)  \,,
\end{aligned}\ee
are a basis of polarization vectors and $\ell\equiv \sqrt{k_2^2+k_3^2}$. We also write $\omega\equiv -k_0$ and $k\equiv k_1$. In these coordinates (not to be confused with the null basis employed above) we take $x^0$ to be the time direction (i.e.\ $\omega$ is indeed a frequency), $x^1$ the radial direction, and $x^2,x^3$ the angular directions. Notice however that we have made no additional symmetry assumptions at this stage, so the terms `radial' and `angular' are used here in anticipation of the axisymmetric spacetimes to be studied below. We also remark that, in a spherical symmetric system, $Z^{(2)}_\mu$ and $Z^{(3)}_\mu$ are respectively even and odd under parity transformations; with some abuse of terminology, we will refer to the corresponding modes as `polar' and `axial', respectively.

Recall that the structure of the Lagrangian \eqref{eq:lagrangian} ensures the existence of precisely three degrees of freedom in the vector field, two transverse and one longitudinal mode. These two sets of modes are decoupled at linear order and thus lead to independent dispersion relations (see Appendix~\ref{app:dispersion relations} for derivations).

For the longitudinal mode we find
\begin{align} \label{eq:longitudinal dispersion}
\omega^2_L = k^2 + \frac{X_2}{X_1}\ell^2 + \frac{X_2}{F_2}  \,,\quad
\mathcal R_L = \frac{\ell^2}{X_1} + \frac{1}{F_2} \,,
\end{align}
where $\mathcal R_L$ corresponds to the norm of the mode, defined as the residue of the mode's propagator at the pole $\omega^2=\omega^2_L$.

The dispersion relations and norms of the polar $(+)$ and axial $(-)$ modes are given by
\begin{widetext}
\be\begin{aligned} \label{eq:polar-axial dispersion}
\omega_\pm^2 &= k^2 + \frac{Y_1 \ell^2+ Y_2}{2F_1F_2}+ \frac{F_2 \ell^2 + X_1}{F_1} \mp \frac{\sqrt{\left(Y_1 \ell^2 + Y_2 \right)^2 + 4 F_2 F_4^2 \ell^2 \left(F_2 \ell^2+ X_1\right)}}{2F_1F_2} \,,\\
\mathcal{R}_\pm &= \frac{1}{F_2 \ell^2 + X_1} k^2 + \frac{1}{2F_1} + \frac{1}{2F_2} \mp \frac{\left( \frac{1}{F_2 \ell^2 + X_1} k^2 + \frac{1}{2F_1} - \frac{1}{2F_2} \right) \left( Y_1 \ell^2 + Y_2 \right) - \frac{F_4^2}{F_1} \ell^2}{\sqrt{\left( Y_1 \ell^2 + Y_2\right)^2 + 4 F_2    F_4^2 \ell^2 \left(F_2 \ell^2 + X_1\right)}} \,,
\end{aligned}\ee
\end{widetext}
where $Y_1 \equiv F_1 F_3 - F_2^2 + F_4^2$ and $Y_2 \equiv F_1 X_2 - F_2 X_1$ depend on spacetime but not on the wavenumbers $k$ and $\ell$.

\section{No-go theorem} \label{sec:no-go}

Stability of the system requires that, for all physical modes, $\omega^2\geq0$ and $\mathcal R>0$ everywhere in spacetime (restricted here to the domain outside the event horizon) and for all values of $k$ and $\ell$. The condition on the sign of $\mathcal R$ is what we refer to as the `ghost stability' condition. The positivity of $\omega^2$ leads to two distinct notions of stability. The property of `gradient stability' means that $\omega^2\geq0$ for large enough values of $k$ and $\ell$; more precisely, the limits $\lim_{k\to\infty}\omega^2/k^2$ and $\lim_{\ell\to\infty}\omega^2/\ell^2$ must be non-negative (and real). The property of `tachyonic stability' means that the quantity $m^2_{\rm eff}\equiv \omega^2\big|_{k,\ell=0}$, interpreted as the effective mass squared of the mode, must be non-negative (and real).

Starting with the longitudinal mode, from \eqref{eq:longitudinal dispersion} we readily deduce
\begin{itemize}[leftmargin = *]
\item Gradient stability:
$X_2 / X_1 \geq 0$;
\item Ghost stability:
$X_1 > 0$ and $F_2 > 0$;
\item Tachyonic stability:
$X_2 / F_2 \geq 0$.
\end{itemize}
Clearly, in this case, the gradient and ghost stability conditions imply the tachyonic stability condition.

The analysis of the polar and axial modes, Eq.~\eqref{eq:polar-axial dispersion}, is significantly more involved but the final results may be stated very compactly:
\begin{itemize}[leftmargin=*]
\item Gradient stability:
$F_1 F_2 > 0$ and $F_1 F_3 \geq 0$;
\item Ghost stability:
$F_1>0$, $F_2 > 0$ and $X_1 \geq 0$;\footnote{Taken independently, the condition of ghost stability is actually more complicated than the indicated inequalities. Our simplified result follows upon applying the gradient condition.}
\item Tachyonic stability:
$X_1 / F_1 \geq 0$ and $X_2 / F_2 \geq 0$.
\end{itemize}
Considered independently, tachyonic stability does not follow from the other two conditions for this sector. However, if we take into account the gradient and ghost stability conditions for \textit{all} modes, we obtain
\be\begin{gathered}
    F_1 > 0\,, \quad F_2 > 0 \,,\quad  F_3 \geq 0 \,,\\
    X_1 > 0 \,,\quad X_2 \geq 0 \,,
\end{gathered}\ee
at each spacetime point. The condition of tachyonic stability for all modes then follows. This is our no-go theorem for spontaneous vectorization.

In fact, we can establish a stronger result: ghost and gradient stability not only implies tachyon stability but also stability for \textit{all} values of $k$ and $\ell$. In other words, our theorem rules out exotic situations where an instability may exist for some intermediate range of momenta, while maintaining a positive $m^2_{\rm eff}$.

\section{Applications} \label{sec:applications}

Having ruled out the possibility of purely tachyonic instabilities, we may next explore the implications of our stability criteria for black hole solutions in GR. As mentioned, the class of Petrov type D spacetimes with diagonal $G^{\mu}{}_\nu$ in the NP basis encompasses most of the well-known stationary black hole metrics, including the Kerr-Newman (with and without cosmological constant), Taub-NUT, accelerated black hole and Bertotti-Robinson solutions; see e.g.~\cite{Astorino:2024bfl} for further examples.

All these cases are actually instances of the so-called Pleba\'{n}ski-Demia\'{n}ski (PD) metric, defined by the conditions of stationarity, axisymmetry, and the existence of two geodesic and shear-free null basis vectors~\cite{Plebanski:1976gy,Griffiths:2005qp,NAKAJIMA2025170030}.\footnote{Notice that the PD metric is contained in the Petrov type D class with diagonal $G^{\mu}{}_\nu$ in the NP basis. We are not aware of any black hole solution in the latter class which is \textit{not} of the PD type, although examples of horizonless solutions do exist.} Since the PD metric is general enough for our purposes here, we restrict our attention to it in this section.

In Boyer-Lindquist-like gauge the PD metric reads
\begin{align}
\Omega^2 ds^2 &= \rho^{-2}( a^2 s^2 h - \Delta ) dt^2 + \rho^2 \Delta^{-1} dr^2 + \rho^2 h^{-1} d\theta^2  \nonumber\\
&\quad +\rho^{-2} \left( (r^2 + a^2)^2 h - \Delta a^2 s^2 \right) s^2 d\phi^2  \nonumber\\
&\quad -2 a s^2 \rho^{-2} \left( (r^2 + a^2) h - \Delta \right) dt d\phi  \,,\label{eq: g_GPD}
\end{align}
where $\rho\equiv\sqrt{r^2 + a^2 x^2}$, and $s,x$ are short-hand notations for $\sin(\theta)$ and $\cos(\theta)$, respectively. $\Omega(rx)$, $\Delta(r)$ and $h(x)$ are arbitrary functions of the indicated arguments, taken to be smooth and positive-definite everywhere in the domain of interest. The latter is precisely defined as the spacetime region where $\Delta>0$, corresponding to the exterior of the (outer) event horizon in the case of a black hole.\footnote{This assumes that the spacetime may be extended beyond the surface $\Delta=0$; otherwise $\Delta=0$ represents a singularity. Although we will not consider examples of this kind, they are also encompassed by our general results. Notice also that we allow $\Delta$ to have another zero beyond the event horizon, as with the cosmological horizon in the case of Kerr-de Sitter.}

We next consider a few specific cases of physical interest. We refer the reader to Appendix \ref{app:PD metric formulae} for some explicit intermediate formulae and to Appendix \ref{app:taub-nut} for another example.

\subsection{Kerr-Newman spacetime}

The Kerr-Newman (KN) spacetime is the unique asymptotically flat black hole solution endowed with spin and electric charge~\cite{Chrusciel:2012jk}. It corresponds to the choice
\be\begin{aligned}  \label{eq: KN}
\Delta = r^2 - 2 M r + M^2 a^2 + Q^2  \,,\quad
h = \Omega = 1  \,,
\end{aligned}\ee
in Eq.~\eqref{eq: g_GPD}, where $M$, $a$ and $Q$ are respectively the mass, spin parameter and charge of the hole. We recall that the outer horizon is located at $r_+ \equiv M + M \sqrt{1 - a^2 -  Q^2/M^2}$ in this coordinate system.

\begin{figure}
\centering
\includegraphics[width=0.35\textwidth]{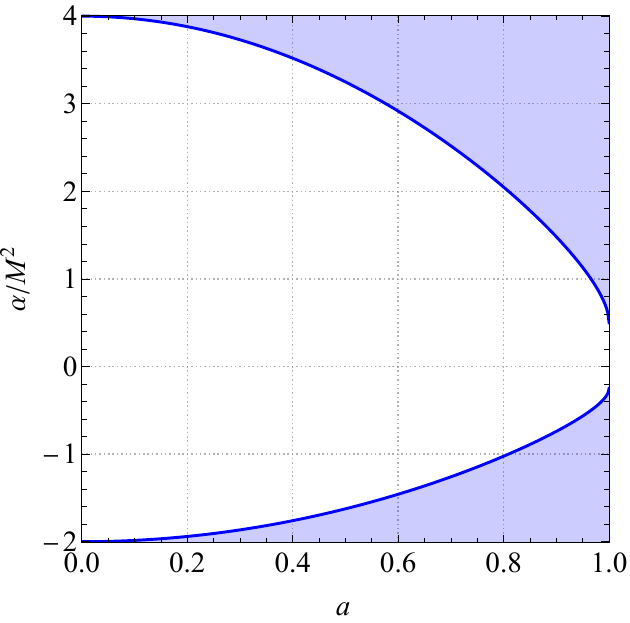}
\caption{Stability region for a Kerr black hole of mass $M$ as function of the spin parameter $a$. Colored region corresponds to instability under ghosts and/or gradients.}
\label{fig:Kerr-G6}
\end{figure}

In the setting of a Kerr black hole ($Q = 0$), the stability conditions only restrict the coupling $\alpha$ since the spacetime is Ricci-flat. Evaluating we obtain
\begin{align}
- \frac{1}{2} < \frac{2M \alpha}{r_{+}^3} < 1  \,.
\end{align}
This condition implies that, for any given non-zero $\alpha$ and fixed $a$, there is a lower bound on the mass $M$ for the black hole to be stable.

Although the same outcome holds in the spinless Schwarzschild case (cf.~\cite{BeltranJimenez:2013btb,Garcia-Saenz:2021uyv}), we find that the stability window narrows as the spin increases, being about an order of magnitude more stringent in the extremal Kerr spacetime ($a=1$); cf.\ Fig.~\ref{fig:Kerr-G6}. More interestingly, the analysis of the rotating system reveals the possibility of dynamical evolution from stable to unstable, since spin may increase in realistic situations, for instance as a result of merger events. Put differently, our results imply that there exists an upper bound on the spin of the black hole whenever $\alpha/M^2<-1/4$ or $\alpha/M^2>1/2$.

\begin{figure*}
    \centering
    \includegraphics[width=0.3\textwidth]{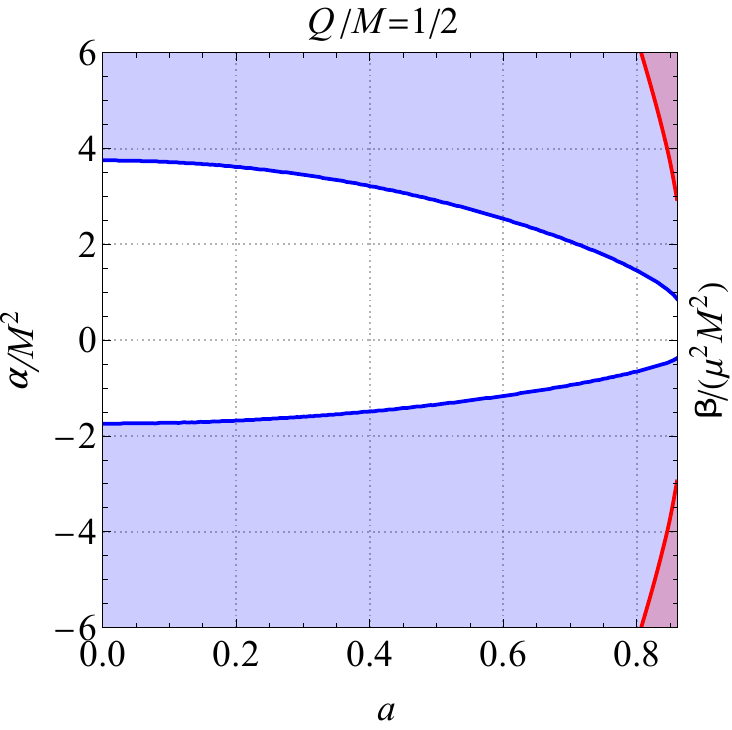}~~~~
    \includegraphics[width=0.3\textwidth]{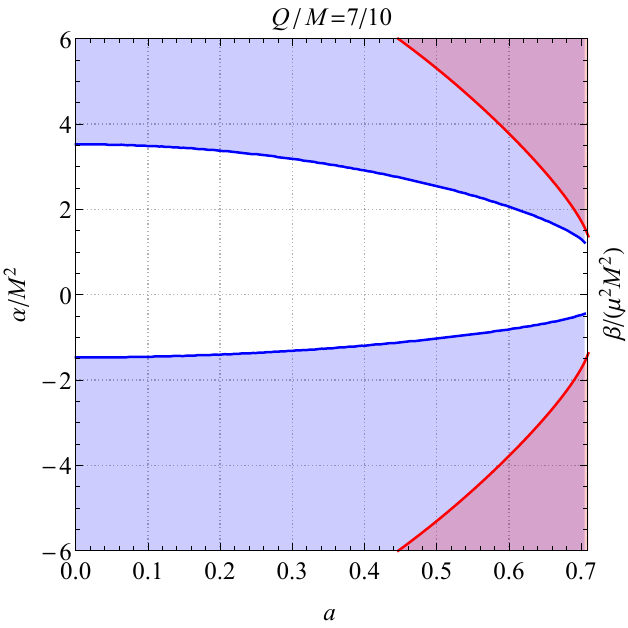}~~~~
    \includegraphics[width=0.3\textwidth]{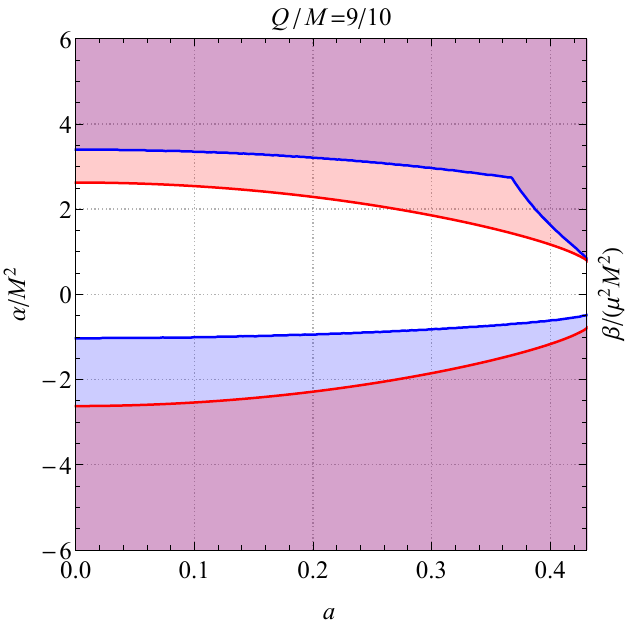}
    \caption{Stability region for a Kerr-Newman black hole. Graphs show instability regions for the couplings $\alpha$ (blue) and $\beta$ (red), given fixed $M$ and $\mu$, as functions of the spin $a$. The three panels correspond to different values of $Q/M$.} \label{fig:KN}
\end{figure*}

The case of a charged black hole is interesting because it has a non-vanishing Einstein tensor, leading also to a constraint on the non-minimal coupling constant $\beta$, more precisely on the combination $\beta/\mu^2$. Since $\beta$ is dimensionless, it may be expected to be of order unity on naturalness grounds, so that the conditions of stability will translate into \textit{lower} bounds on the vector particle mass $\mu$. The explicit result is
\begin{align}
-1 < \frac{2\beta Q^2}{\mu^2 r_+^4} \leq 1  \,.
\end{align}
We illustrate the stability bounds in Fig.~\ref{fig:KN}. It is clear that, for fixed non-zero charge $Q$, the window of stability for $\beta$ (or $\mu$) features a strong dependence on the black hole spin, especially in the regime of small charge of most astrophysical interest (cf.~\cite{Levin:2018mzg,Gong:2019aqa,Dai:2019pgx,Chen:2021sya,Adari:2021qmx} for estimates of values of charge that may be achieved in astrophysical situations).

\begin{figure}
    \centering
    \includegraphics[width=0.35\textwidth]{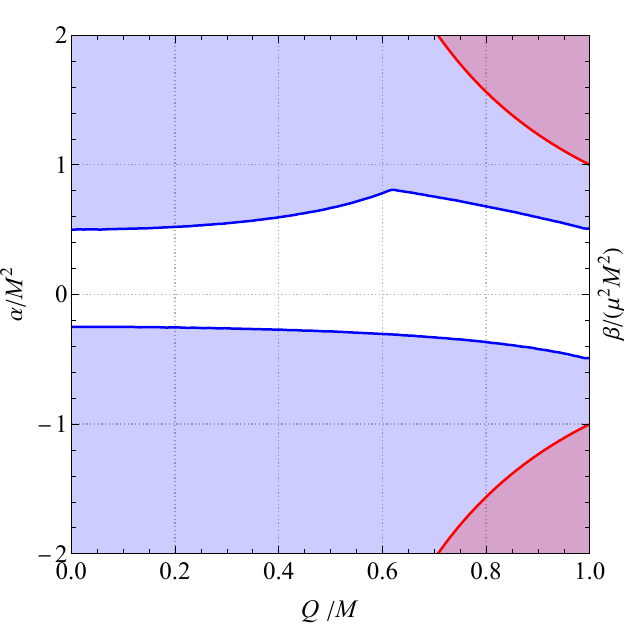}
    \caption{Stability region for an extremal Kerr-Newman black hole. Graph shows instability regions for the couplings $\alpha$ (blue) and $\beta$ (red), given fixed $M$ and $\mu$, as functions of the charge-to-mass ratio $Q/M$.} \label{fig:extreme}
\end{figure}

It is algebraically more involved but straightforward to derive explicit expressions for the stability bounds on $\alpha$. For simplicity we display here only the results for the extremal case ($r_+=M$), where we fix $a=\sqrt{1-Q^2/M^2}$ for given $Q/M$:
\be\begin{aligned}
- \frac{1}{4-2Q^2/M^2} &< \frac{\alpha}{r_+^2} < \frac{1}{2-2Q^2/M^2} \,,\\
\text{and}\qquad \frac{\alpha}{r_+^2} &\leq 1 - \frac{Q^2}{2M^2}  \,,
\end{aligned}\ee
where the latter inequality applies only if $Q^2/M^2 > \frac{3-\sqrt{5}}{2}$. The resulting stability regions are shown in Fig.\ \ref{fig:extreme}.

\subsection{Accelerated black hole spacetime}

As a second example we consider the so-called C-metric solution of GR~\cite{EhlersKundt1962}. It describes a pair of accelerating black holes connected by a line defect, representing a simple set-up to study the effects of acceleration on black hole horizons. For us, it serves as a toy model to assess how acceleration may affect black hole stability in theories with non-minimally coupled vector fields.

The C-metric is described by the line element \eqref{eq: g_GPD} with $a=0$ and the choice~\cite{Griffiths:2005se,Griffiths:2006tk}
\be\begin{aligned}
\Delta &= \left(1-A^2 r^2/r_h^2\right) \left(r^2-r_h r\right)  \,,\\
h &= 1-Ax  \,,\quad
\Omega^2 = 1-Ax r/r_h \,,
\end{aligned}\ee
where $r_h = 2M$ is the Schwarzschild radius and $A > 0$ is a dimensionless parameter that quantifies the acceleration of the black hole (note that the second black hole is not visible in this coordinate system). The surface $r=r_h$ corresponds to an event horizon provided $0<A<1$. There is also a so-called `acceleration horizon' located at $r=r_h/A$, analogous to the Rindler horizon for a uniformly accelerated observer in flat spacetime. The spacetime exterior to the event horizon corresponds to the domain $r_h<r<r_h/A$.

\begin{figure}
    \centering
    \includegraphics[width=0.35\textwidth]{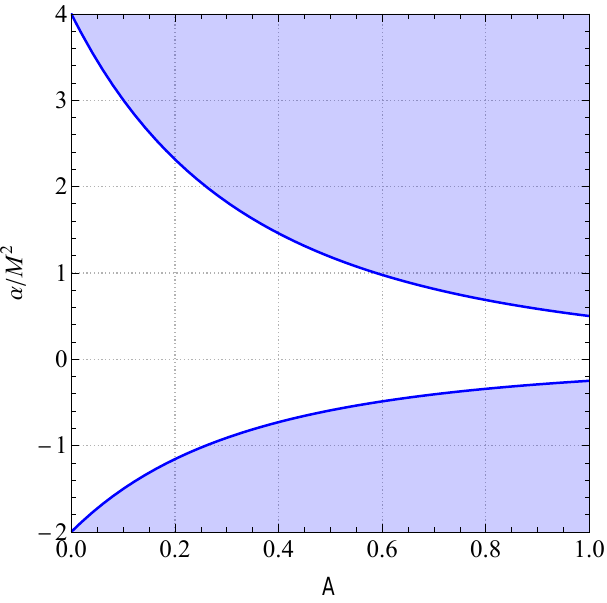}
    \caption{Stability region for an accelerated black hole of mass $M$ described by the C-metric, as function of the acceleration parameter $A$. Colored region corresponds to instability under ghosts and/or gradients.} \label{fig:acceleration}
\end{figure}

Evaluating the stability criteria we obtain
\be
- \frac{1}{2} < \frac{2M \alpha}{r_h^3}\left(1+A\right)^3 < 1 \,.
\ee
The bounds are displayed in Fig.\ \ref{fig:acceleration}. In qualitative similarity to the Kerr black hole case, we see that acceleration also results in a narrowing of the stability window and the possibility of transiting into the unstable region via dynamical evolution.

\subsection{Kerr-(anti-)de Sitter spacetime}

Also interesting is to study the effects of a cosmological constant on the stability criteria, having in mind potential applications to holographic systems (see e.g.~\cite{Jing:2010zp,Jing:2010cx}). We focus here on the Kerr-(anti-) de Sitter metric, although the inclusion of electric charge would be a straightforward generalization.

The line element is again given by \eqref{eq: g_GPD} with
\be\begin{aligned}
\Omega^2 \Delta &= ( r^2 + M^2 a^2 ) ( 1 - r^2 \Lambda ) - 2 M r  \,,\\
\Omega^2 h &= 1 + a^2 r^2 \Lambda  \,,\quad
\Omega = \left( 1 + M^2 a^2 \Lambda \right)^{1/4}  \,,
\end{aligned}\ee
where $\Lambda$ is the cosmological constant, $\Lambda>0$ for de Sitter (dS) and $\Lambda<0$ for anti-de Sitter (AdS). The spacetime may exhibit up to three distinct horizons; whenever any of these coincide, we speak of a (doubly or triply) degenerate point in parameter space.

Although somewhat cumbersome, it is straightforward to analyze the stability conditions. Since there are effectively two black hole free parameters in this case ($a$ and $\Lambda$ upon normalizing everything by the hole's mass $M$), we choose for simplicity to consider two particular values of the spin $a$ and derive the stability bounds as functions of $\Lambda$.

\begin{figure}
\centering
\includegraphics[width=0.75\linewidth]{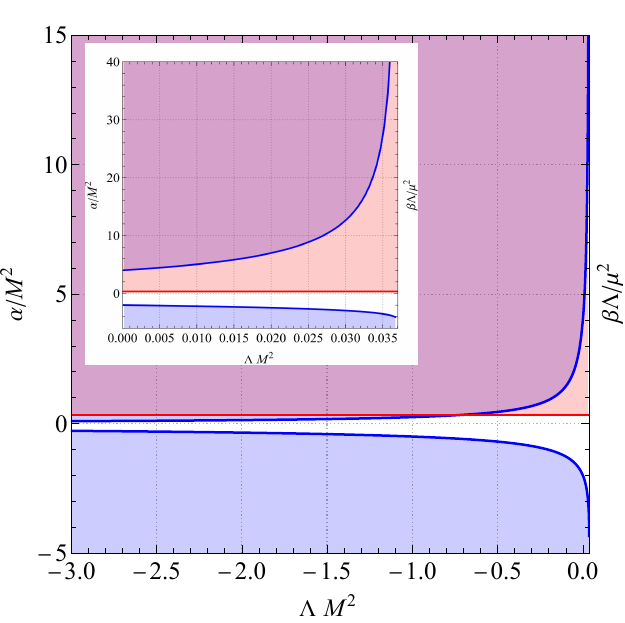}
\caption{Stability region for a Schwarzschild-(A)dS black hole. Graph shows instability regions for the couplings $\alpha$ (blue) and $\beta$ (red) as functions of the cosmological constant $\Lambda$. The maximum value of $\Lambda$ corresponds to the degenerate dS point. The inset is a zoomed-in version of the graph restricted to positive $\Lambda$.}
\label{fig:SdS}
\end{figure}

Consider first $a=0$, i.e.\ the case of Schwarzschild-(A)dS. The stability criteria then reduce to
\be\begin{aligned}
&-\frac{1}{2}<\frac{\alpha}{r_h^2}<\frac{1}{1-3\Lambda r_h^2} \,,
\\ &\text{and}\qquad  3\beta \Lambda <\mu^2 \,,
\end{aligned}\ee
where $r_h$ is the location of the event horizon, i.e.\ the smaller positive root of $f=1-2M/r-r^2\Lambda$. Notice that $r_h>0$ requires $\Lambda M^2\leq \frac{1}{27}$, otherwise the spacetime describes a naked singularity. The stability domain is displayed in Fig.~\ref{fig:SdS}.

Next we consider the extremal case where the two innermost horizons coincide, and we denote their common radius $r_h$. We then find concise expressions for the stability conditions on the non-minimal coupling constants:
\be\begin{aligned}
\frac{1}{a^2 - 3r_h/M} &< \frac{2 M^2\alpha}{r_h^4\Omega^2} < \frac{1}{a^2} \,,\\
\text{and}\qquad 3\beta\Lambda &< \mu^2 \Omega^2 \,,
\end{aligned}\ee
where $a$ and $r_h$ are fixed in terms of $M$ and $\Lambda$ by the condition of extremality. Real solutions however only exist for a range of values of $\Lambda$, explicitly $-\frac{64}{27} \leq \Lambda M^2 \leq \frac{16}{(3 + 2 \sqrt{3} )^3}$, with the saturation of the inequalities corresponding to the triply degenerate AdS and dS points, respectively.

\begin{figure}
\centering
\includegraphics[width=0.75\linewidth]{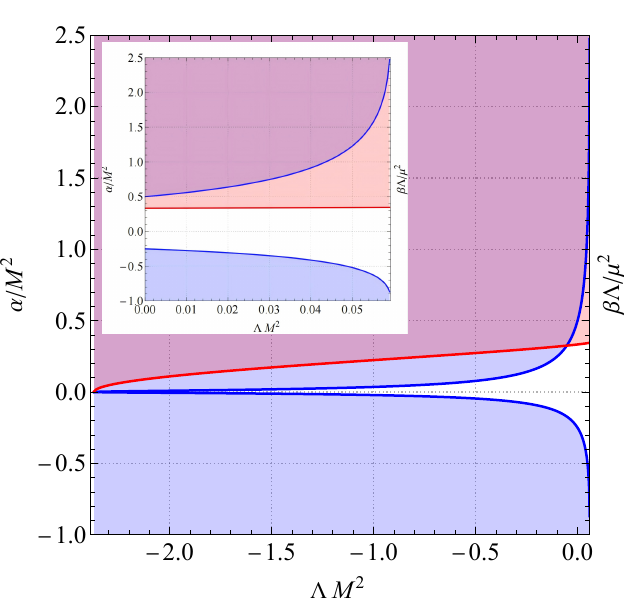}
\caption{Stability region for an extremal Kerr-(A)dS black hole with degenerate (inner) horizons. Graph shows instability regions for the couplings $\alpha$ (blue) and $\beta$ (red), given fixed $M$ and $\mu$, as functions of the cosmological constant $\Lambda$. The minimum and maximum values of $\Lambda$ correspond to the triply degenerate AdS and dS points, respectively. The inset is a zoomed-in version of the graph restricted to positive $\Lambda$.}
\label{fig:KdS}
\end{figure}

The resulting bounds are shown in Fig.~\ref{fig:KdS}. Interestingly, the stability window for $\alpha$ closes at the triply degenerate AdS point. It becomes wider for larger $\Lambda$ but does not fully open at the triply degenerate dS point, signaling an asymmetry between dS and AdS. Furthermore, in contrast to Kerr-Newman, the bound on $\beta$ for Kerr-(A)dS is single-sided. This can be understood in terms of the equation of state of matter in each case. If we define $w=G^\mathbf{34}/G^\mathbf{12}$ (i.e.\ the ratio of the tangential pressure to the energy density) as a rough measure of the local equation of state parameter, then the stability bound on $\beta$ may be written (locally) as $-1 < 2\beta G^\mathbf{12}/\mu^2 \leq w^{-1}$. For Kerr-Newman, i.e.\ with electromagnetic energy, we have $w=1$, leading to a double-sided bound. On the other hand, for a cosmological constant $w= -1$, hence the single-sided bound. This argument also suggests that in the more realistic situation of a Kerr black hole surrounded by ordinary matter (i.e.\ $0<w<1$) one would have asymmetric bounds on $\beta$. It would be interesting to confirm this for some explicit examples.

\section{Discussion}

It is worth summarizing the assumptions of our no-go theorem, which may suggest potential loopholes or paths to generalizations or alternative mechanisms for the generation of black hole vector hair.

First, we have studied the linear stability of hairless solutions. Our theorem does not rule out the possibility of a transition to a hairy state via non-linear dynamics. Relatedly, our criterion that ghost- and gradient-type instabilities are inadmissible may be unwarranted, for instance if non-linear effects were to quench the instabilities (see e.g.~\cite{Bonanno:2025ohx}).

Second, the set-up we have considered includes two non-minimal vector-tensor couplings. As explained, these terms are unique in that they do not introduce additional degrees of freedom. On the other hand, one may adopt the viewpoint of effective field theory and declare that such extra modes lie outside the domain of validity of the theory. This would open the door to more general set-ups, which may also be tackled with our methods.

Third, while the class of spacetimes to which our theorem applies is very broad, it is still characterized by a high degree of symmetry. One may envisage the possibility that astrophysical black holes ``start life'' with vector hair, i.e.\ that hair generation is concomitant with the dynamical processes leading to the formation of black holes. Such processes are of course highly non-symmetric and thus not captured by our analysis.

Fourth, our theorem relies on the eikonal approximation. While this limit must correctly capture ghost- and gradient-type instabilities, it may in principle misdiagnose tachyon-type instabilities. It has been shown that for a Schwarzschild black hole the eikonal analysis is in fact accurate~\cite{Garcia-Saenz:2022wsl}, but we see no reason why the same conclusion should apply to more general systems.

Even though our main interest and motivation stems from the question of black hole hair formation, it is worth emphasizing that our main results make no assumption about the existence of an event horizon. Thus, our theorem in principle applies equally well to other astrophysical objects or systems of theoretical interest. Notice however that rotating perfect fluid solutions, as would be relevant for stellar models, have a less symmetric Einstein tensor than the one assumed here and are therefore not encompassed by our set-up (although solutions of Petrov type D do exist~\cite{Stephani:2003tm}). Nevertheless, the fact that the theorem does apply in the case of non-rotating fluids~\cite{Garcia-Saenz:2021uyv,Silva:2021jya} suggests that a similar result should persist upon including spin. We believe that these considerations motivate the extension of our analysis to more general metrics.

\textit{Acknowledgments.---}This work received support from the NSFC (Grant No.\ 12505074) and from a Provincial Grant (Grant No.\ 2023QN10X389).

\appendix

\section{Dispersion relations} \label{app:dispersion relations}

In this Appendix we provide some details of our derivations of the dispersion relations and mode residues, Eqs.\ \eqref{eq:longitudinal dispersion} and \eqref{eq:polar-axial dispersion}. For the general formalism, we refer the reader to Ref.~\cite{Garcia-Saenz:2021uyv}.

The vector field equation of motion that derives from \eqref{eq:lagrangian} reads
\be
4\mathcal G^{\mu\nu\rho\sigma}\nabla_\mu F_{\rho\sigma} + 2\mathcal V^{\mu\nu}A_\mu=0 \,.
\ee
We substitute the mode decomposition $A_\mu=\sum_n C_n Z^{(n)}_\mu e^{iS}$ and keep the terms quadratic in $k_{\mu}\equiv \partial_{\mu}S$ in addition to the mass matrix $\mathcal V^{\mu\nu}$,
\be\begin{aligned}
0&=\left( 2 \mathcal G^{\mu\nu\rho\sigma} k_\mu k_\rho  + \mathcal V^{\nu\sigma} \right) Z^{(n)}_\sigma C_n + \ldots \\
&=\sum_{m,n} (Z^{-1})^{\nu}_{(m)} \Delta^{-1}_{mn} C_n +\ldots \,,
\end{aligned}\ee
where the last equality defines the propagator matrix $\Delta$. Here $Z^{-1}$ is the inverse of the polarization basis introduced in~\eqref{eq:Z_basis} and the ellipses stand for subleading terms in the eikonal approximation. Notice that we assume a hierarchy in which the mass matrix dominates over other terms and is therefore not neglected.

The theory \eqref{eq:lagrangian} enjoys a Lorenz constraint,
\be\begin{aligned}
0 &= \nabla_\mu ( \mathcal V^{\mu\nu}  A_\nu ) \\
&\propto X_1 (-\omega C_0 + k C_1 ) + X_2 \ell C_2 + \ldots \,,
\end{aligned}\ee
which provides a natural projection of the momentum $k^\mu$ onto the vector field space. More precisely, we may identify the polarization vector of the longitudinal mode, $Z^{(L)}_\mu$, as the solution of the equation of motion that lies in the subspace spanned by the wave-vector $k_{\mu}$ and the direction defined by the Lorenz constraint, i.e.\ $L_\mu\equiv X_1 (-\omega  Z^{(0)} + k  Z^{(1)} ) + X_2 \ell  Z^{(2)}$. Explicitly we find
\be
Z^{(L)} \propto ( -F_2 \omega \ell, F_2 k \ell, F_2 \ell^2 + X_1 , 0 ) \,.
\ee
Given this, we introduce a new basis of the field space $Z^{(\tilde n)} \equiv \left( Z^{(0)}, Z^{(1)}, Z^{(L)}, Z^{(3)} \right)$, and observe that the propagator matrix is a tensor under field reparametrizations, i.e.
\be
\Delta^{-1}_{\tilde m \tilde n} = \sum_{m,n} \left( \frac{\delta Z^{(\tilde m)}}{\delta Z^{(m)}} \right)^{-1}\left( \frac{\delta Z^{(\tilde n)}}{\delta Z^{(n)}} \right)^{-1}\Delta^{-1}_{mn} \,,
\ee
in this new basis. The advantage is that $\Delta^{-1}$ (and hence $\Delta$) is now manifestly diagonal in the longitudinal mode, as claimed. Thus the equation ${\rm det}\,\Delta^{-1}=0$ that determines the mode spectrum has a solution given by $\Delta^{-1}_{LL}=0$, with residue $\mathcal R_L\equiv -\lim_{\omega^2\to\omega^2_L}(\omega^2-\omega_L^2)\Delta_{LL}$ (the sign is conventional), leading to the results quoted in \eqref{eq:longitudinal dispersion}.

The remaining components of $\Delta^{-1}_{\tilde m \tilde n}$ read
\be\begin{aligned}
\Delta^{-1}_{00} &= F_1 k^2 + F_2 \ell^2 + X_1  \,,\\
\Delta^{-1}_{11} &= F_1 \omega^2 - F_2 \ell^2 - X_1  \,,\\
\Delta^{-1}_{33} &= F_2 (\omega^2 - k^2 ) - F_3 \ell^2 - X_2  \,,\\
\Delta^{-1}_{01} &= F_1 \omega k  \,,\quad
\Delta^{-1}_{03} = F_4 k \ell  \,,\quad
\Delta^{-1}_{13} = F_4 \omega \ell  \,.\\
\end{aligned}\ee
A more involved but direct calculation then yields the dispersion relations and residues for the polar and axial modes, as given explicitly in the main text, Eq.\ \eqref{eq:polar-axial dispersion}.

\section{Applications to the Pleba\'{n}ski-Demia\'{n}ski metric} \label{app:PD metric formulae}

For reference, we list here the explicit expressions for the functions $F_i$ and $X_i$ for the general PD metric, Eq.\ \eqref{eq: g_GPD}:
\be\begin{aligned}
F_1 &= 1 - \alpha \Omega^2 \rho^{-6} \big{(} 2 \left( r^2 - 3 a^2 x^2 \right) \left( \Delta - z^2 h \right) \\
&\quad - 4 z^2 \rho^2 x h' + \rho^4 s^2 h'' \big{)}  \,,  \nonumber\\
F_2 &= 1 + \alpha \Omega^2 \rho^{-6} \big{(} 2 ( r^2 - a^2 x^2 ) \Delta \\
&\quad + 2 a^2 \big{(} \rho^2 - 2 r^2 s^2 \big{)} h - \rho^2 \left(r \Delta' + a^2 s^2 x h'\right) \big{)}  \,, \\
F_3 &= 1 - \alpha \Omega^2 \rho^{-6} \big{(} 2 \left( 3 r^2 - a^2 x^2\right) \left( \Delta - a^2 s^2 h \right) \\
&\quad- 4 \rho ^2 r \Delta' + \rho ^4 \Delta '' \big{)}  \,, \\
F_4 &= - 3 \alpha \Omega^2 a x \rho^{-6} \big{(} 4 r \Delta - 2 r \left(z^2 + a^2 s^2\right) h \\
&\quad- \rho^2 \left( \Delta' - r s^2 x^{-1} h' \right) \big{)} \,,\\
X_1 &= \mu^2 + \frac{\beta}{\alpha} (3 - 2 F_2 - F_1)  \,,\\
X_2 &= \mu^2 + \frac{\beta}{\alpha} (3 - 2 F_2 - F_3)  \,,
\end{aligned}\ee
where $z^2 \equiv r^2+a^2$ and here primes denote derivatives with respect to $r$ and $x$ respectively for $\Delta$ and $h$.

To obtain the above expressions starting from the definitions in \eqref{eq:F defs} one needs the Newman-Penrose (NP) basis vectors in the Boyer-Lindquist coordinate system used in this paper. It is easier to first obtain the Cartesian basis vectors,
\be\begin{aligned}
e^0{}_\mu &= \Omega^{-1} \rho^{-1} \Delta^{1/2} \left(1,0,0,-a s^2\right)  \,,\\
e^1{}_\mu &= \Omega^{-1} \rho \Delta^{-1/2} \left(0,1,0,0\right)  \,,\\
e^2{}_\mu &= \Omega^{-1} \rho h^{-1/2} \left(0,0,1,0\right)  \,,\\
e^3{}_\mu &= \Omega^{-1} \rho^{-1} h^{1/2} s \left(-a,0,0,r^2+a^2\right) \,,
\end{aligned}\ee
which by definition satisfy $e^a{}_{\mu}e^b{}_{\nu}\eta_{ab}=g_{\mu\nu}$. The NP basis vectors $\{\ell,n,m,\bar{m}\}$ are then given in terms of the Cartesian ones by (as a particular choice)
\be\begin{aligned}
\ell&= \frac{1}{\sqrt{2}}\left(e^0+e^1\right) \,,\\
n&= \frac{1}{\sqrt{2}}\left(e^0-e^1\right) \,,\\
m&= \frac{1}{\sqrt{2}}\left(e^2+ie^3\right) \,,\\
\bar{m}&= \frac{1}{\sqrt{2}}\left(e^2-ie^3\right) \,.
\end{aligned}\ee

\section{Taub-NUT spacetime} \label{app:taub-nut}

Another well-known example of an axisymmetric black hole solution is the Taub-NUT metric~\cite{Taub:1950ez,Newman:1963yy}. It is of theoretical interest due to its properties of self-duality and the presence of a topological or `NUT' charge, among many other features~\cite{Ortin:2015hya}. The metric also belongs to the PD class considered in the main text.

\begin{figure}
    \centering
    \includegraphics[width=0.35\textwidth]{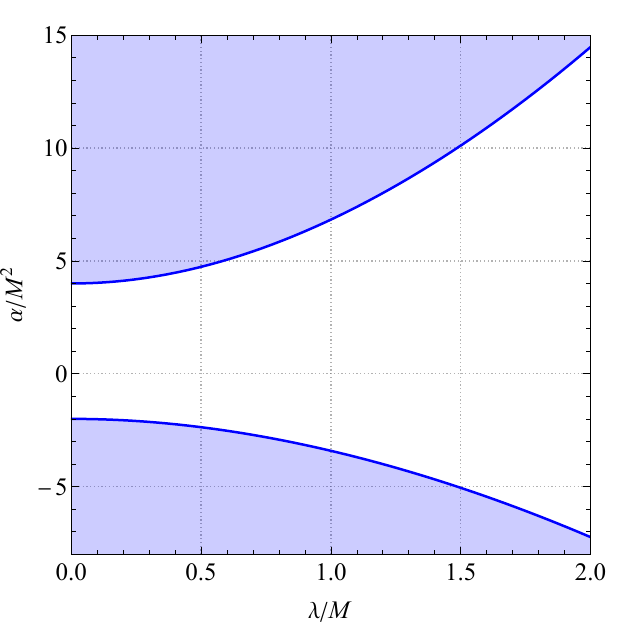}
    \caption{Stability region for a Taub-NUT black hole of mass $M$ as function of the NUT charge $\lambda$. Colored region corresponds to instability under ghosts and/or gradients.} \label{fig:NUT}
\end{figure}

We focus on the electrically neutral Taub-NUT metric for simplicity, which is a vacuum solution of GR, leading to simple stability criteria for the non-minimal coupling $\alpha$,
\be\begin{aligned}
-\frac{1}{2} < \alpha r_+^2 \frac{4\left(M^2+\lambda^2\right)}{\left(r_+^2+\lambda^2\right)^3} < 1 \,,
\end{aligned}\ee
where $\lambda$ is the NUT charge and $r_+\equiv M+\sqrt{M^2+\lambda^2}$ is the outer horizon. The domain of stability is also shown in Fig.\ \ref{fig:NUT}. We see that for this system the bounds are tightest when $\lambda=0$ (i.e.\ Schwarzschild black hole), and become wider as $\lambda$ increases. This may be understood from the fact that the curvature scales as $1/\lambda^2$, so that the geometry becomes flat, in a local sense, in the limit of large $\lambda$.

\bibliography{main.bib}

\end{document}